\documentstyle[emulateapj,flushrt,eqsecnum]{article}

\begin{document}
\def\fg{$f_{ICM}$ }
\def\RI{$R_I$ }
\def\LRI{$\log{R_I}$ }
\def\TX{$\left<T_X\right>$ }
\def\LTX{$\log{\left<T_X\right>}$ }
\def\ST{ST }
\def\beff{$\beta_{eff}$ }
\def\LT{$L_X-\left<T_X\right>$ }
\def \rfiveh {\hbox{$r_{500}$}}
\def \etal      {{\it et al.\ }}
\def\spose#1{\hbox to 0pt{#1\hss}}
\def\lta{\mathrel{\spose{\lower 3pt\hbox{$\mathchar"218$}}
     \raise 2.0pt\hbox{$\mathchar"13C$}}}
\def\gta{\mathrel{\spose{\lower 3pt\hbox{$\mathchar"218$}}
     \raise 2.0pt\hbox{$\mathchar"13E$}}}

\submitted{The Astrophysical Journal: submitted May 9, accepted July 11, 1997}

\title{An X--ray Size--Temperature Relation for Galaxy Clusters:\\
Observation and Simulation}

\author{Joseph J. Mohr\altaffilmark{1,2} \& August E. Evrard\altaffilmark{1,3}}

\affil{$^1$Department of Physics \& $^2$Department of Astronomy, University of Michigan, Ann Arbor, MI 48109 \\ 
$^3$Institut d'Astrophysique, 98bis Blvd Arago, 75014 Paris, France}

\authoremail{jjmohr@umich.edu}\authoremail{evrard@umich.edu}

\begin{abstract}

We show that galaxy clusters conform to a tight
relation between X--ray isophotal size \RI and emission weighted
intracluster medium (ICM) temperature \TX.
The best fit relation for 41 members of an X--ray flux limited cluster sample
is:
$\log{R_I}=\ (0.93\pm0.11)\ \log{(\left<T_X\right>/6\,{\rm keV})}-\ 
(0.08\pm0.01)$;
intrinsic scatter in size about the relation is $15\%$,
and for 30 clusters with 
$\left<T_X\right>>4$~keV the scatter is reduced to $10\%$.  
The existence of the size--temperature (ST) relation indicates the ICM
structure is a well behaved function of $\left<T_X\right>$.
We use an ensemble of gasdynamic 
simulations to demonstrate that a cluster population
experiencing present epoch growth nonetheless conforms to an
\ST relation with scatter similar to that observed;  the simulations also
exhibit a tight relation between $M_{vir}$ and \TX, providing the 
suggestion that a similar relation holds for observed clusters.
We use the scatter in \RI to estimate limits on the {\it rms} 
variation in ICM mass fraction $\delta f_{ICM}$ at constant \TX: 
$\delta f/f_{ICM}\le22$\% ($\le14$\% for clusters 
with $\left<T_X\right>>4$~keV).  It appears that a mechanism like
feedback from galaxy winds, which introduces systematic structural changes
in the ICM, is required to reproduce the observed slope of the \ST relation.  

\end{abstract}


\section{INTRODUCTION}

Ongoing cluster growth at the present epoch is an observational fact
(e.g. Fabricant et al. 1986, 1989, \cite{zabludoff95,henry95,mohr96});
large studies employing varied techniques indicate that 
30--70\% of clusters exhibit evidence for recent mergers
(\cite{geller82,dressler88,jones92,mohr95}).
A theoretical framework for understanding this growth has emerged, 
and structure formation simulations within a range of favored
cosmological models produce clusters which exhibit merger signatures
similar in detail and frequency to those observed
(\cite{mohr95,buote97}).

Yet, recent numerical work indicates that even though clusters are
growing at the present epoch, they are, on average, regular objects with
temperatures and masses tightly correlated through the virial theorem 
(\cite{evrard96,schindler96,rbl96}). 
Direct observational tests of these results are difficult, but
analyzing relations between ICM structural parameters and temperature 
provides an alternative means of testing cluster regularity.
The well known X--ray luminosity--temperature (\LT)
relation (e.g. \cite{smith79,mitchell79,david93,mushotzky97})
is one observational indicator of cluster
regularity, but the scatter around the relation is quite large.
The scatter is largely due to varying degrees of excess core emission 
associated with cooling flows (Fabian \etal 1994).  

Here we report a tight relation between cluster
isophotal size \RI and mean, emission weighted temperature \TX 
that we uncovered in an ongoing
morphological analysis of a large, X--ray flux limited cluster sample.
This \ST relation for observed clusters indicates a high degree of regularity,
suggests the existence of a mass--temperature correlation, and provides
new constraints on theories of the interactions between galaxies and
the ICM. This newly discovered \ST relation may ultimately lead to 
more accurate estimates of cluster baryon fractions and 
tighter constraints on the cosmic density parameter 
$\Omega_0$ (e.g. \cite{white93,evrard97}).

Section 2 contains a description of the data reduction, analysis and the
observed \ST relation.  We then use an ensemble of numerical simulations to
examine the \ST relation in a cluster population experiencing present epoch
growth ($\S3$).  In Section 4, we discuss sources of scatter about
the relation, estimate upper limits on the variation in ICM mass fraction
\fg, and discuss the slope of the \ST relation.  Section 5 contains
a discussion of our conclusions.

\section{OBSERVATIONS}

The cluster sample contains members of the X--ray flux limited group of
55 clusters defined by Edge \etal (1990).
Archival ROSAT Position Sensitive Proportional
Counter (PSPC) images of 47 of these
clusters are available online through the High Energy Astrophysics Science
Archive Research Center (HEASARC).  We reduce these images
using PROS and Snowden analysis software
(\cite{snow94}).  The final image for each observation is the sum of
individually flatfielded
subimages from the Snowden bands R4 through R7, corresponding to photon
energies 0.44--2.04~keV.  We exclude time intervals with master
veto rates higher than 220~cts/s,  and exclude other high
background time intervals (typically $\le$5~min excluded) whose
inclusion would degrade the detection significance of a source
10\% as bright as the background (\cite{pildis95}).  We combine
multiple images of a cluster using the positions of bright X--ray point
sources where possible, or alternatively, the image header pointing positions.
Finally, we remove obvious point sources from the unsmoothed
images (we do not remove possible point source components coincident with the
extended emission typical of cluster cooling flows, as in NGC~1275, for
example) and then Gaussian smooth ($\sigma=2$~pixels). 
The pixel scale is $14''.947$, and the effective angular resolution
is $FWHM\sim1.5'$.  We correct the cluster images for $(1+z)^4$
cosmological dimming, but do not make galactic absorption or $K$ corrections
(discussed in more detail below).

We define the cluster size \RI using the area $A_I$ of the largest region
enclosed by the isophote $I$. 
\begin{equation}
R_I= \sqrt{A_I/\pi}
\end{equation}
This approach deals consistently with clusters that are not
azimuthally symmetric because of recent mergers, and produces a 
tighter relation than would azimuthal averaging;  in addition, 
$I$ can be chosen so that \RI is unaffected by the
central surface brightness excesses typical of cooling flows.
The isophote $I$ is the cluster surface
brightness, and we determine the X--ray background $I_B$ using annuli well
outside the X--ray bright cluster region (typically
38$'$ from the cluster center with $\sim8'$ extent).  
We estimate $I_B$ as the
(sigma clipped) mean value of the pixels within the annulus,
and we use the width of the distribution around $I_B$ to estimate the
uncertainty.  For the isophotes $I$ we use, the uncertainty in
\RI is primarily due to the background uncertainty. 
We conservatively estimate the uncertainties in \RI to be
$\sigma_R=(R_{I_-} - R_{I_+})/2$ where $I_{+/-}=I\pm I_B/10$.
The internal estimate of the uncertainty in $I_B$ is significantly smaller
than 10\% in all cases.

We use published, emission weighted, mean temperatures \TX for each
cluster.  There are 42 clusters in the Edge sample with PSPC images and
published \TX with uncertainties.
Roughly half of these temperatures are from {\it Einstein} MPC observations
(\cite{david93}), but where possible we substitute more accurate
{\it Ginga}, {\it ASCA}, or ROSAT PSPC temperatures
({\it Ginga}--- \cite{day91,allen92,johnstone92,david93,hughes93,arnaud97};
{\it ASCA}--- \cite{henrik96,marke96,matsu96,tamura96,marke97};
PSPC--- \cite{david96}).
We exclude the Virgo cluster from our analysis because bright, cluster
emission extends beyond the 2$^\circ$~PSPC field of view.
Thus, our final sample contains 41 clusters.  

The observational data for
this sample are displayed in Table \ref{datatable}; the cluster name,
effective PSPC exposure time $t_{exp}$, background brightness $I_B$,
emission weighted mean temperature \TX, and designation as cooling flow
cluster are listed.  We designate those clusters
with central cooling times significantly below 10~Gyr (\cite{edge92})
as cooling flow clusters.

\begin{figure}
\plotfiddle{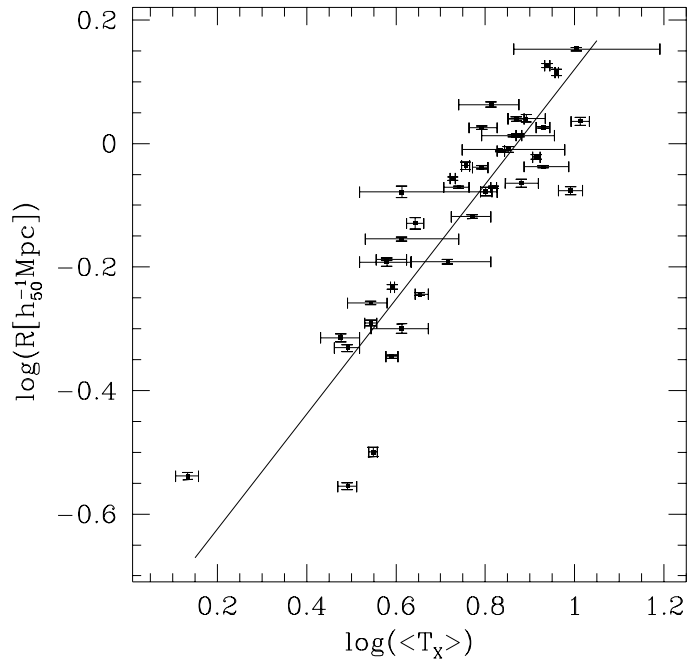}{3.3in}{0}{90}{90}{-190}{-150}
\caption{The X--ray \ST relation for 41 clusters.
The cluster redshift is used to calculate \RI, and
$I=1.93\times10^{-3}$~cts/s/arcmin$^2$; error bars are $1\sigma$. 
The best fit \ST relation
(see Eqn \ref{bestfit}) is plotted, and the
{\it rms} scatter around this relation is $\delta\log{R_I}=0.084$.
The intrinsic scatter about the relation (removing contribution from
\TX uncertainties) is $\delta\log{R_I}=0.064$ (15\%); for 30 clusters with
$\left<T_X\right>>4$~keV, the scatter is $\delta\log{R_I}=0.041$ (10\%).
}\label{scaleobs}
\end{figure}

Figure \ref{scaleobs} contains a plot of the 41 clusters in
\LRI and \LTX for $I=1.93\times10^{-3}$~cts/s/arcmin$^2$.  A linear
fit to the sample (minimizing the sum of the orthogonal residuals from
the relation) yields
\begin{equation}
\log{R_I}=\ (0.93\pm0.11)\ \log{{\left<T_X\right>\over6\,{\rm keV}}}-\ 
(0.08\pm0.01).
\label{bestfit}
\end{equation}
The {\it rms} scatter in \LRI about this relation is
$\delta\log{R_{raw}}=0.084$. The coefficient uncertainties
reflect the half width of the 68\% confidence region
determined by bootstrap resampling.

We analyze the \ST relation in the PSPC sample over a factor of 8 in surface
brightness (see Table \ref{fittable} and $\S$3.2);
varying $I$ primarily changes the zeropoint $b$ of the \ST relation.
The integrity of the relation is preserved
until $I$ reaches values comparable to the peak surface brightnesses
of some clusters.

The uncertainties in \TX are large enough to contribute to the observed
scatter.  We estimate the intrinsic scatter in $\delta\log{R_I}$ 
by subtracting the temperature contribution in quadrature.  We find
$\Delta(\delta\log{R_I})=0.93*RMS(\sigma_T/T_X/\ln{10})=0.054$.
Thus, the intrinsic scatter is approximately $\delta\log{R_{int}}=0.064$.
Fig.~\ref{scaleobs} gives the impression that the scatter increases at
lower temperature; fitting only those 30 clusters 
with $\left<T_X\right>>4$~keV leads to an \ST relation
$\log{R_I}=\,(0.82\pm0.14)\,\log{(\left<T_X\right>/6\,{\rm keV})}
-\,(0.07\pm0.02)$, with
raw/intrinsic scatter of $0.067/0.041$.  Thus, the slope of the 
\ST relation for hot clusters is slightly shallower than that for the 
whole population, and the scatter is smaller by $50\%$.

\section{SIMULATIONS}

We further examine the \ST relation using a set of
48 high resolution N--body plus gas dynamics simulations, to be described
in detail elsewhere (\cite{mohr97b}).  In summary, the cluster
simulations are carried out within four different cold dark matter 
(CDM) dominated cosmologies 
(1) SCDM ($\Omega=1$, $\sigma_8=0.6$, $h=0.5$),
(2) OCDM ($\Omega_0=0.3$, $\sigma_8=1.0$, $h=0.8$),
(3) LCDM ($\Omega_0=0.3$, $\lambda_0=0.7$, $\sigma_8=1.0$, $h=0.8$),
and (4) ZCDM ($\Omega=1$, $\sigma_8=1.0$, $h=0.5$) with power spectrum
consistent with a CDM $\Gamma=0.24$ transfer function (\cite{davis85}).  
Here $H_0=100h$~km/s/Mpc and 
$\sigma_8$ is the power spectrum normalization in $8 h^{-1}$ Mpc spheres. 
The baryon density is a fixed fraction of the total 
$\Omega_b = 0.2 \Omega_0$.  Within each of these models, we use two $128^3$
N--body only simulations of cubic regions with scale $\sim400$~Mpc to
determine sites of cluster formation.  Within these initial runs the
virial regions of clusters with Coma--like masses of $10^{15}$~M$_\odot$
contain $\sim$10$^3$ particles.

Using the N--body results for each cosmological model,
we choose clusters for additional study.
We zoom in on these clusters, resimulating them at higher resolution with
gas dynamics and gravity on a $64^3$ N--body grid.  The simulation scheme is
P3MSPH (\cite{evrard88}), and radiative cooling is ignored. 
These clusters have masses which varying by an order of magnitude.
The scale of the 
simulated region surrounding each cluster are in the range 50--100~Mpc, and
vary as $M_{halo}^{1/3}$, where $M_{halo}$ is approximately the mass enclosed
within the present epoch turn around radius.  Thus, the 48 simulated clusters
in our final sample have similar fractional mass resolution; the spatial
resolution varies from 125--250~kpc.  We create X--ray images and temperature
maps for further analysis following procedures described in Evrard (1990).

Fig.~\ref{scalesims}A is a plot of the \ST relation for
the simulated clusters (with $I=1.62\times10^{-3}h^3_{50}$~cts/s/arcmin$^2$).
Clusters from the four cosmological models
are plotted with different point styles, each appears three times 
from orthogonal projections.  The best fit
relation for the SCDM model is
$\log{R_I}=0.70\log{(\left<T_X\right>/6\,{\rm keV})} + 0.119$
and the scatter around this fit is $\delta\log{R}=0.042$;
Table \ref{fittable} contains the relations and scatter for all models; 
the level of scatter is typically somewhat smaller than the observed
value.  Only ZCDM clusters have a higher scatter than our estimate
of the intrinsic scatter in the observed sample; the higher scatter is
caused by a few ongoing major mergers at high $\left<T_X\right>$.  

The scatter around the \ST relation in the simulated populations
is due to projections of clusters along the line of sight, small
equilibrium departures, and small ($\sim5$\%) intrinsic fractional 
\fg variations which exist because of different formation histories.
On average, these clusters reached 
half of their present epoch mass at look--back epochs 
0.33$t_0$: 0.35$t_0$: 0.42$t_0$: 0.44$t_0$: for Z:S:O:LCMD, where $t_0$ is
the age of the universe in each model.  
The S, O and LCDM clusters are morphologically similar to observed clusters,
and the ZCDM clusters exhibit somewhat more evidence for recent mergers
than the other three models and the observed clusters (\cite{mohr97b}).

\section{ANALYSIS}

The existence of the \ST relation in this large, approximately flux
limited cluster sample demonstrates that the ICM structure is a
well behaved function of $\left<T_X\right>$. After correcting for measurement 
uncertainties, the scatter of the clusters about the \ST relation is
similar to the scatter of bright ellipticals around the fundamental
plane (e.g., \cite{jor96,mohr97}).  To further understand this
relation, we examine sources of scatter about it, and then we use the
scatter to place upper limits on variations in the ICM mass fraction \fg.
We also examine the \ST relation slope, which provides information about
the physical processes affecting the ICM structure.

\subsection{Scatter About the \ST Relation}

There are several sources of scatter in the \ST relation 
besides the measurement
uncertainties discussed in $\S2$.  These include: (1) variations in galactic
$N_H$ absorption, (2) departures from equilibrium, (3) projections of
physically unassociated clusters along the line of sight, and
(4) dark matter and ICM structural variations at constant $\left<T_X\right>$.

The contribution from variations in galactic $N_H$ is insignificant.
We use PROS tasks to
calculate surface brightness correction factors for galactic absorption at
1~keV for our cluster sample.  We find that these correction factors have
a mean of 1.17 and an {\it rms} variation of 6.8\% about this mean value for
the 41 clusters. Typical surface brightness profiles are rather steep 
$I(R)\propto R^{-3}$ (e.g. \cite{jones84,mohr95}), 
so $\delta \log{R} = {1\over3} \delta\log{I}$ and the variation in 
galactic absorption along the lines of sight to these clusters is
expected to contribute scatter at the level of $\sim$2.3\%.  Galactic
absorption is approximately 45\% more effective at 0.5~keV,
the low energy end of our bandwidth
(and, of course, less effective at 2.0~keV),
so the maximal contribution to the \ST relation scatter
from variations in galactic $N_H$ is $\sim$3.3\%.

Departures from equilibrium contribute significantly to the scatter.  Two
clusters of similar mass about to merge in the plane of the sky will
be 40\% too large in \RI for their $\left<T_X\right>$.
The scatter about the \ST relations exhibited by the simulations described
in $\S3$ is an indicator of the expected contribution from equilibrium
departures and projection effects in cluster populations
still merging at the present epoch.

Equilibrium dark matter and ICM structural variations at a particular
temperature will also contribute to the \ST relation scatter.  N--body
simulations indicate that the dark matter structure of clusters is
a regular function of mass, and that mild variations among clusters with
similar mass are attributable to different formation times (\cite{navarro96}).
Again, major mergers present exceptions, but do not dominate the local 
population.  The ICM distribution of clusters with similar masses has not 
been as extensively studied as the dark matter, but the experiments shown 
here and below suggest that the ICM structure is a well behaved 
function of cluster mass or temperature, even in the presence of 
galactic winds (Metzler \& Evrard 1997).  Other sources of variation 
in the ICM distribution or gas mass fraction are possible.  
One source is radiative cooling, which 
leads to strong emission excesses or cooling flows in
cluster centers (\cite{jones84,fab94}).  

We examine the effects of cooling flows by dividing our cluster sample into
those with cooling times ($t_{cool}$ significantly less than 10~Gyr)
short enough that cooling flows should exist in the core,
and those with longer cooling times (\cite{edge92}).  We find that the
best fit \ST relation for the 18 cooling flow clusters
(see Table \ref{datatable}) has a slope
$m=0.93\pm0.16$ and a zeropoint $b=-0.07\pm0.02$;
the relation for the 23 non--cooling flow clusters has
$m=1.00\pm0.21$ and $b=-0.09\pm0.02$.  
The raw/intrinsic scatter about these two
relationships is $\delta\log{R_I}=0.086/0.071$ for the cooling
flow clusters and $\delta\log{R_I}=0.083/0.061$ for the non--cooling
flow clusters.  These two relations are
indistinguishable and the scatter about each is comparable;  thus,
for the present sample of 41 clusters there is no indication that the
\ST relation is sensitive to the presence of cooling flows.

\subsection{Upper Limit on ICM Mass Fraction Variations}

By assuming the observed scatter in the \ST relation is caused
solely by varying ICM mass fraction \fg at constant \TX,
we estimate an upper limit on these variations.
The surface brightness profiles of both real and simulated clusters 
are generally well fit with the standard, isothermal $\beta$--model 
\begin{equation}
I(R) = A\Lambda\rho_0^2 R_c\left[1+(R/R_c)^2\right]^{-3\beta+1/2}
\label{betaeq}
\end{equation}
where $\rho_0$ is the central ICM density, $\Lambda$ is the specific
emissivity, $R_c$ is a core radius, $\beta$ is the variable slope, and
$A$ is a constant with some $\beta$ dependence (\cite{cav78}).
We use this model as a guide to understanding the \ST relation (but
the relationship below can be derived independent of the $\beta$ model).
Within this framework, the ICM mass fraction is
\begin{equation}
f_{ICM} = {4\pi\rho_0 R_c^3 \int_0^{R_{vir}/R_c}d\lambda\,\lambda^2
(1+\lambda^2)^{-3\beta/2}\over M_{vir}}
\label{gmass}
\end{equation}
where $R_{vir}$ is the virial radius encompassing a 
fixed density contrast.  
Outside the core, the surface brightness scales as 
\begin{equation}
I(R) \ = \ C \ f_{ICM}^2 \ R^{-6\beta +1} 
\label{betaeq2}
\end{equation}
where the constant $C$ incorporates the ICM structural information 
of equation~(\ref{gmass}).  If clusters of a given temperature 
are structurally similar, then 
the scatter in the isophotal radii at constant \TX places 
an upper limit on variations in the gas fraction $\delta\log{f_{ICM}}
\le [(6\beta-1)/2]\delta\log{R_I}$.  This inequality can be broken if 
structural variations (variations in $C$) are anti--correlated with
variations in $f_{ICM}$; we assume here that this is not the case.  
Qualitatively, such an anti--correlation would require gas--poor
clusters to be more centrally concentrated (and vice--versa), a
situation at odds with traditional expectations.  

For the typical value $\beta = 2/3$ 
(\cite{jones84,mohr95}), $\delta\log{f_{ICM}}\le1.5\delta\log{R_I}$, 
and the intrinsic scatter in \LRI limits fractional gas  
variations to $\delta f/f_{ICM} \le 22\%$.  The limit drops to 
$\delta f/f_{ICM} \le 14\%$ for the 30 clusters hotter than 4 keV.  

The scatter about the \ST relation in the simulated clusters
includes a contribution from variations in \fg within the virial 
radius.  This contribution is rather small: 
$\delta f/f_{ICM} \simeq 4\%$ within $\rfiveh$, 
the radius within which the mean 
cluster density is $500$ times the critical density.  
Thus, other effects such as departures from equilibrium and projections
dominate the scatter.  
Assuming these effects exist to a similar degree in 
the observations, then the similar scatter in the simulations and
the observed clusters with $\left<T_X\right>>4$~keV 
indicates that 
gas fraction variations within hot clusters are very small, perhaps 
as small as $4\%$.  This argument does not apply for cooler clusters, 
which exhibit larger scatter about the \ST relation.

\subsection{Slope of the \ST Relation}

The steeper observed slope $m=0.93\pm0.11$, compared to
$0.61\le m\le0.81$ in the simulations, suggests structural 
differences in real clusters compared to this set of models.   
Radiative cooling is not included in the simulations, but the
similarity of the \ST relation in clusters with and without
cooling flows indicates this is not the cause of the slope differences.
Possible explanations include systematic trends in the 
overall gas fraction \fg and/or systematic variations in the 
ICM distribution (as measured by $\beta$) with $\left<T_X\right>$.

The observed cluster sample provides evidence that shallower
surface brightness profiles lead to steeper \ST relations.
Table \ref{fittable} lists the best fit slope $m$ and zeropoint $b$ for
the observed clusters over a factor of 8 variation in $I$.  Along with
the expected zeropoint decrease with brightening $I$, there is a tendency
for a steeper slope.  This steepening is related to
the shallower surface brightness profiles (smaller effective $\beta$) at
brighter isophotes.  Simulations from the four cosmological models provide
additional evidence that shallower surface brightness profiles generally
lead to steeper \ST relation slopes; the mean $\beta$'s of the 
Z:S:L:OCDM are are 0.67:0.73:0.77:0.81 and the $m$'s 
are 0.81:0.70:0.72:0.61.  
Thus, the steeper observed \ST relation may simply be an indication of
a more extended ICM distribution in the observations than in the simulations.

Galactic winds are one physical process expected to introduce systematic ICM
structural variations.
In such a model, star formation within galaxies results in
supernovae (SNe) driven winds which expell gas, and the expelled
gas has the orbital kinetic energy of its parent galaxy plus the energy
imparted by the SNe.  This non--gravitational source of energy is expected
to be more important in low \TX clusters where the orbital energies are
smaller.  We examine the effects of galactic 
winds on the \ST relation using an ensemble of clusters from simulations
which include galaxy feedback (Metzler \& Evrard 1994; 1997).  The simulated
cluster ensemble contains two populations; one population of 18 clusters
evolved consistent with an SCDM model and another population with the same
initial conditions and cosmological parameters, but with galaxy feedback
modeled by discrete gas ejection totalling half the initial galaxy mass
between $z=4.5$ and the present.  This extreme ejection model is
intended to estimate the maximal effect of galaxy feedback.

\begin{figure}
\plotfiddle{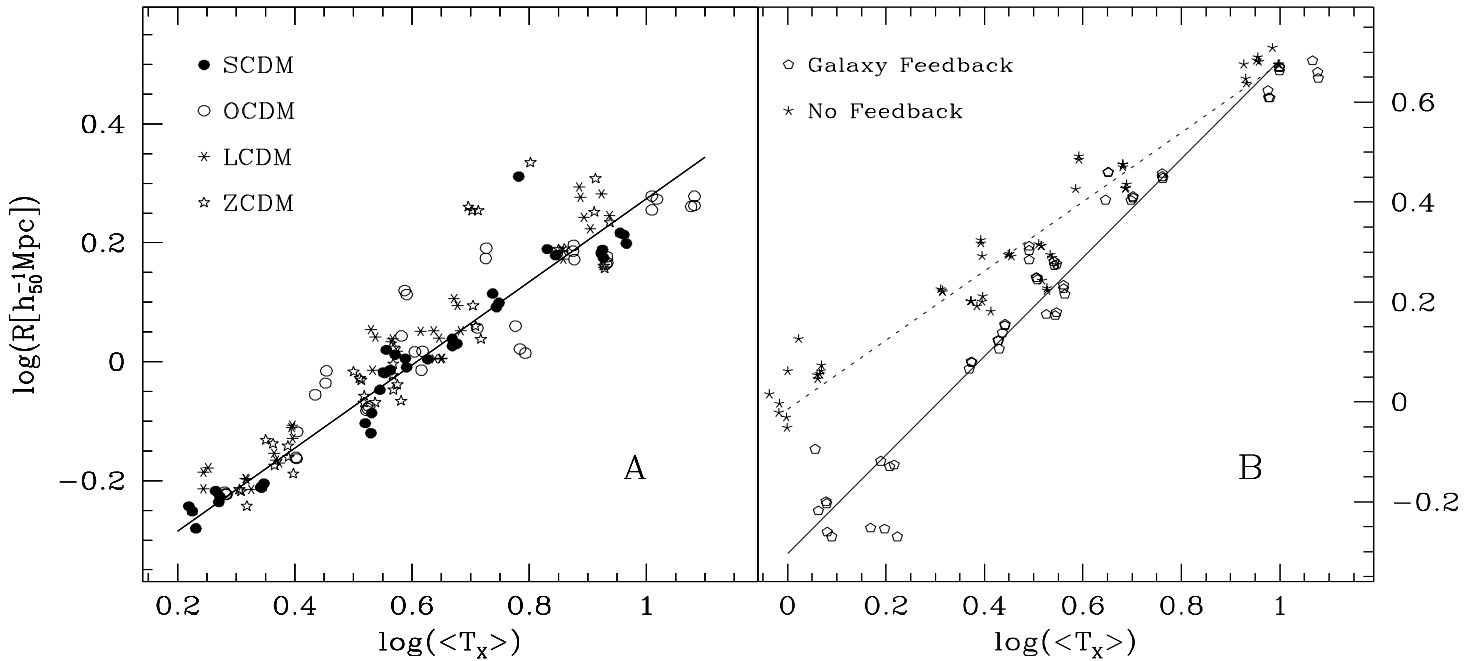}{3.0in}{0}{100}{90}{-310}{-150}
\caption{The \ST relations for simulated galaxy clusters.
Our 48 simulations appear on the left (A), with the
points coded according to cosmological model. 
The best fit relation for the SCDM model is plotted (see Table \ref{fittable}),
and the {\it rms} scatter around this relation is $\delta\log{R}=0.042$. 
The scatter for these simulated clusters
(which have formed recently) is similar to that for observed clusters, but
the slope of the relation is shallower.  On the right (B) we demonstrate
the effects of galaxy feedback on the \ST relation
with simulations from Metzler \& Evrard (1994, 1997).
The best fit slope for the clusters simulated with feedback ($m=0.99$)
is much steeper than the slope of those simulated without feedback ($m=0.69$).
}\label{scalesims}
\end{figure}

Fig.~\ref{scalesims}B is a plot of the effects of 
galaxy feedback on the \ST relation.  The slope
of the best fit \ST relation for the clusters with no feedback
(dotted line) is $m=0.69$, consistent with the SCDM slope in our
simulations (see Table \ref{fittable}).  The slope 
for clusters simulated with ejection (solid line) is
$m=0.99$, somewhat steeper than the slope of our PSPC sample.  
This demonstrates that galaxy feedback can cause the structural changes
required to match the slope of the observed \ST relation.  
Structural changes include shallower surface brightness profiles
(smaller $\beta$) and lower overall gas fractions with 
decreasing $\left<T_X\right>$, as discussed in 
Metzler \& Evrard (1997).
Fig.~\ref{scalesims}B makes clear that, as expected, ejection has the
greatest effect on the lowest mass clusters. 

The tendency for low \TX clusters to have shallower surface brightness
profiles is well known (e.g., \cite{mohr95});
our sample of 41 clusters exhibits this general trend. 
Fig.~\ref{betafig} contains
a plot of the slope of the surface brightness profile \beff
measured over the region of the cluster used to calculate \RI.
The Spearman rank correlation coefficient (\cite{press92}) for \TX and \beff
is $r_s=0.54$ with a correlation significance of 99.97\%. 
\beff is a two dimensional generalization of $\beta$,
and like \RI, is measured without azimuthal averaging; 
moreover, \beff reflects the local slope of
the surface brightness profile and so is independent of the cluster core
radius.  Briefly, to calculate the values in Fig.~\ref{betafig} we
determine the area $A_l$ enclosed by the isophote
$I_l=1.29\times10^{-3}$~cts/s/arcmin$^2$ (chosen to be somewhat fainter than
the $I$ used to calculate \RI),
and then calculate the isophote $I_h$ which encloses the area $A_h$ which
is chosen to be $A_l/2$. 
This provides two isophotes and two areas, and so we then calculate
the effective slope of the surface brightness profile between $I_l$ and $I_h$
\begin{equation}
\beta_{eff}={1\over3}\left({\log{\left(I_h/I_l\right)} \over 
\log{\left(A_l/A_h\right)}}+{1\over2}\right)
\end{equation}
This form follows directly from the expression $I(R)\propto R^{-6\beta_{eff}+1}$.
The trend of falling \beff with \TX is qualitatively consistent with the
effects of galaxy feedback.

\section{DISCUSSION}

We use observations of the 41 members of an X--ray flux limited
cluster sample with PSPC observations and published \TX (except Virgo)
to demonstrate that nearby clusters conform to a tight relation
between X--ray isophotal size and cluster temperature. 
The intrinsic scatter in size at fixed temperature is 
only $15\%$ ($10\%$ for clusters with $\left<T_X\right>>4$~keV). 
The existence of the \ST relation indicates that the ICM structure 
outside the core regions is a well
behaved function of \TX and suggests a tight correlation between 
\TX and $M_{vir}$ as seen in our simulations and others
(\cite{evrard96,schindler96,rbl96}).

The \ST relation is significantly tighter than the \LT relation for 
this cluster sample.  The scatter (corrected for temperature uncertainties)
in 2--10~keV X--ray luminosity 
(\cite{david93}) around the best fit relation for these
41 clusters is $\delta L_X/L_X=52$\%.  (We use the same set of
temperature measurements in both cases.)  The factor of 3.5 larger
scatter in the \LT relation reflects the sensitivity of $L_X$ to
cluster core properties (\cite{fabian94}).  When restricted to 
weak cooling flow clusters --- those with inferred cooling flow rates 
below 100 M$_\odot$ yr$^{-1}$ --- the \LT relation displays a much 
smaller dispersion, and inferred gas fraction variations from a 
sample of 24 clusters are 
consistent with the limits set by this analysis (Arnaud \& Evrard
1997).  The narrow scatter in the \ST relation indicates that 
structural regularity exists outside the core at observationally 
viable surface brightness levels. 

Our cluster sample provides no evidence that cooling flows significantly
affect the \ST relation.  The 13 clusters with central cooling times
significantly below 10~Gyr exhibit a best fit slope,
zeropoint, and scatter which are statistically indistinguishable from those
for the 18 other clusters.  The zeropoint differences
between these two \ST relations indicate that \TX in
cooling flow clusters is biased low by $\sim5\pm7$\%, consistent
with previous estimates of $\sim$10--20\% (\cite{fabian94}).

The scatter around the \ST relation is comparable to that
of the half light radius of
elliptical galaxies around the fundamental plane (e.g. \cite{jor96,mohr97});
an obvious use of the \ST relation is as a 
distance indicator.  Knowledge of \TX and an apparent isophotal size 
predicts the cluster distance with an uncertainty limited by the 
scatter in \RI ($\sim10$\% for hot clusters).  
We are currently evaluating the promise of this relation as a
distance indicator at intermediate redshift.

The scatter about the best fit \ST relation for
observed clusters provides an upper limit on the {\it rms}
variation in cluster ICM mass fraction at a common temperature $\left<T_X\right>$. 
Ignoring possible correlations between cluster structural changes 
and \fg variations, we find upper limits $\delta f/f_{ICM}\le22$\%
($\le14$\% for $\left<T_X\right>>4$~keV). 

We use 48 cluster simulations within 4 cosmological models
to demonstrate that cluster populations experiencing growth at the
present epoch exhibit \ST relations with scatter similar to that
observed above $\left<T_X\right>>4$~keV: $\delta{R_I}/R_I \simeq
10\%$.  Only a small contribution to the scatter in the simulations 
is from \fg variations, suggesting that the ICM mass fraction 
in hot clusters may be limited to $\lta 5\%$ variation within their  
virial regions.  Implications for hierarchical models of galaxy 
formation remain to be explored.  

\begin{figure}
\plotfiddle{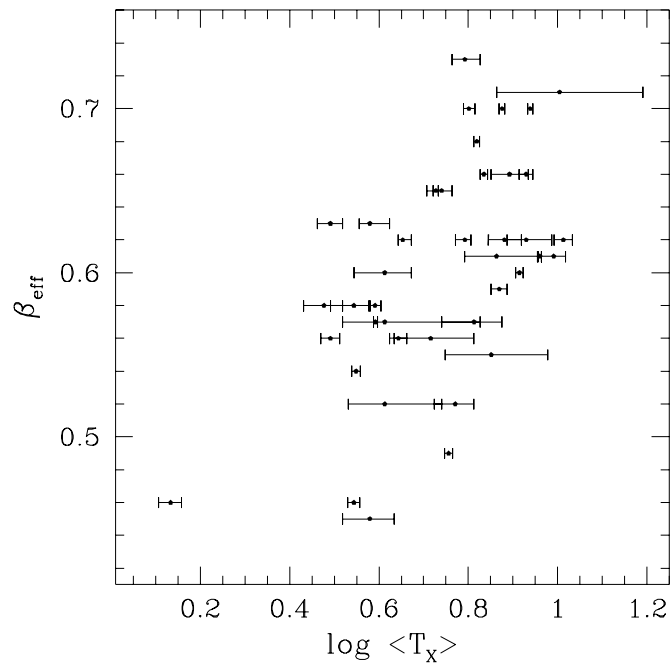}{3.3in}{0}{90}{90}{-190}{-150}
\caption{Slopes of the X--ray surface brightness profiles are plotted
versus cluster temperature $\left<T_X\right>$.  
The trend for cooler clusters to have
shallower profiles is qualitatively consistent with the effects of
galaxy feedback.  \beff is calculated independent of the cluster core
radius and requires no azimuthal averaging (see $\S4.3$ for details).
}\label{betafig}
\end{figure}

The slope of the \ST relation for the PSPC clusters
is steeper than the slope in our simulations.  We use numerical
simulations (\cite{metzler94}, 1997) to show that galaxy feedback
introduces the kinds of ICM structural changes required to steepen the \ST
relation.  The changes include
shallower surface brightness profiles and lower \fg with
decreasing $\left<T_X\right>$. A decrease in $\beta$ with \TX
(see Fig.~\ref{betafig}) and a reduction of \fg in low \TX systems
(\cite{david95,dell95}) have been observed.   Galactic feedback is a promising
mechanism for creating the observed, structural differences between low and
high \TX clusters.

\acknowledgements
We are very grateful to C. Metzler for providing images of his 
simulations, M. Arnaud for providing \TX measurements
prior to publication, M. Geller for comments on the paper, 
and D. Richstone for stimulating discussions.  
We thank the referee for helpful comments.  This research has made use of the
NASA/GSFC HEASARC Online Service, and has been supported by NASA through
grants NAG5--3401 and NAGW--2367.  AEE acknowledges support from the 
CIES and CNRS of France during a sabbatical stay at IAP.

\begin{deluxetable}{lrccrc}
\tablewidth{0pt}
\tablecaption{Galaxy Cluster Data}
\tablehead{
\colhead{Cluster}	&
\colhead{$t_{exp}$\tablenotemark{a}}	&
\colhead{$I_B$\tablenotemark{b}}	&
\colhead{$R_I$\tablenotemark{c}}	&
\colhead{$\left<T_X\right>$\tablenotemark{d}}	&
\colhead{Flow}}
\startdata
A262       &  8.1 & 3.31 & 0.289 &  1.36 & X \nl
MKW3s      &  9.2 & 5.80 & 0.485 &  3.00 & X \nl
A1060      & 14.6 & 5.34 & 0.279 &  3.10 &   \nl
A2052      &  5.9 & 6.06 & 0.467 &  3.10 & X \nl
A1367      & 17.7 & 3.07 & 0.512 &  3.50 &   \nl
A4059      &  5.2 & 2.90 & 0.552 &  3.50 &   \nl
A3526      &  3.0 & 7.09 & 0.317 &  3.54 & X \nl
A3562      & 18.5 & 4.19 & 0.643 &  3.80 &   \nl
A780       & 17.3 & 2.34 & 0.649 &  3.80 & X \nl
AWM7       & 12.6 & 3.34 & 0.452 &  3.90 & X \nl
A496       &  8.1 & 3.82 & 0.586 &  3.91 & X \nl
A2063      &  9.6 & 6.90 & 0.502 &  4.10 & X \nl
A3112      &  7.2 & 2.64 & 0.700 &  4.10 & X \nl
CygA       &  8.7 & 7.07 & 0.835 &  4.10 & X \nl
A2147      &  0.9 & 6.40 & 0.743 &  4.40 &   \nl
A2199      &  9.7 & 2.79 & 0.570 &  4.50 & X \nl
A3391      &  5.5 & 3.24 & 0.644 &  5.20 &   \nl
A1795      & 24.8 & 3.09 & 0.878 &  5.34 & X \nl
A3158      &  2.9 & 2.60 & 0.850 &  5.50 &   \nl
A3558      & 27.8 & 4.75 & 0.920 &  5.70 &   \nl
A119       & 14.3 & 2.82 & 0.761 &  5.90 &   \nl
A3266      &  7.0 & 3.35 & 1.061 &  6.20 &   \nl
A85        &  5.3 & 3.18 & 0.915 &  6.20 & X \nl
A426       &  4.4 & 7.40 & 0.836 &  6.33 & X \nl
A3667      & 11.3 & 4.39 & 1.156 &  6.50 &   \nl
A644       &  9.5 & 2.18 & 0.849 &  6.59 &   \nl
A478       & 21.4 & 1.48 & 0.973 &  6.84 & X \nl
A2244      &  2.9 & 2.31 & 0.978 &  7.10 &   \nl
A2255      & 12.5 & 2.11 & 1.030 &  7.30 &   \nl
A399       &  6.4 & 2.29 & 1.096 &  7.40 &   \nl
A2256      & 17.1 & 2.61 & 1.030 &  7.51 &   \nl
A3571      &  5.5 & 4.98 & 0.862 &  7.60 &   \nl
A2029      & 11.9 & 5.94 & 1.098 &  7.80 & X \nl
A1656      & 19.4 & 5.21 & 0.951 &  8.21 &   \nl
0745-19    &  9.1 & 1.97 & 0.917 &  8.50 & X \nl
A754       &  6.1 & 2.41 & 1.062 &  8.50 &   \nl
A2142      &  4.8 & 2.65 & 1.337 &  8.68 & X \nl
A2319      &  2.9 & 4.92 & 1.303 &  9.12 &   \nl
Ophi       &  3.7 & 6.38 & 0.839 &  9.80 &   \nl
A1689      & 13.3 & 2.53 & 1.422 & 10.10 &   \nl
Tria       &  6.6 & 4.20 & 1.086 & 10.30 &   \nl
\enddata
\tablenotetext{a}{Exposure [ks]}
\tablenotetext{b}{Background [$10^{-4}$ cts/s/arcmin$^2$]}
\tablenotetext{c}{[$h_{50}^{-1}$~Mpc]}
\tablenotetext{d}{[keV]}
\label{datatable}
\end{deluxetable}

\begin{deluxetable}{lrcrcc}
\tablewidth{0pt}
\tablecaption{Fit Results for Size--Temperature Relation}
\tablehead{
\colhead{Source}	&
\colhead{$I$\tablenotemark{a}}	&
\colhead{$m$}	&
\colhead{$b$}	&
\colhead{$\delta\log{R}$\tablenotemark{b}}}
\startdata
PSPC	 & 9.65 & 0.886 (0.124) & -0.030 (0.013) & 0.074\nl
PSPC	 & 19.3 & 0.934 (0.109) & -0.080 (0.013) & 0.064\nl
PSPC	 & 38.6 & 1.047 (0.130) & -0.204 (0.014) & 0.069\nl
PSPC	 & 77.2 & 1.514 (0.382) & -0.356 (0.024) & 0.151\nl
SCDM	 & 16.2 & 0.699 (0.035) & 0.119 (0.013) & 0.042\nl
OCDM	 & 66.4 & 0.612 (0.025) & 0.111 (0.010) & 0.054\nl
LCDM	 & 66.4 & 0.721 (0.024) & 0.156 (0.007) & 0.035\nl
ZCDM	 & 16.2 & 0.808 (0.067) & 0.161 (0.022) & 0.069\nl
\enddata
\tablenotetext{a}{[$10^{-4}$ cts/s/arcmin$^2$]}
\tablenotetext{b}{Intrinsic scatter (corrected for errors in \TX)}
\label{fittable}
\end{deluxetable}

\begin{thebibliography}{}

\bibitem[Allen et al. 1992]{allen92} Allen, S. W., Fabian, A. C., Johnstone, R. M., Nulsen, P. E. J. \& Edge, A. C. 1992, \mnras, 254, 51

\bibitem[Arnaud \& Evrard 1997]{arnaud97} Arnaud, M. \& Evrard, A. E. 1997, in preparation

\bibitem[Buote \& Xu 1997]{buote97} Buote, D. A. \& Xu, G. 1997,
\mnras, 284, 439

\bibitem[Cavaliere \& Fusco--Femiano 1978]{cav78} Cavaliere, A. \&
Fusco--Femiano, R. 1978, A\&A, 70, 677

\bibitem[David et al. 1993]{david93} David, L. P., Slyz, S. C., Forman, W.,
Vrtilek, S. D. \& Arnaud, K. A. 1993, \apj, 412, 479

\bibitem[David, Jones \& Forman 1995]{david95} David, L. P., Jones, C. \&
Forman, W. 1995, \apj, 445, 578

\bibitem[David, Jones \& Forman 1996]{david96} David, L. P., Jones, C. \& Forman, W. 1996, \apj, 473, 692

\bibitem[Davis et al. 1985]{davis85} Davis, M., Efstathiou, G., Frenk, C. S.
\& White, S. D. M. 1985, \apj, 292, 371

\bibitem[Day et al. 1991]{day91} Day, C. S. R., Fabian, A. C., Edge, A. C., Raychaudhury, S. 1991, \mnras, 252, 394

\bibitem[dell'Antonio, Geller \& Fabricant 1995]{dell95} dell'Antonio, I. P.,
Geller, M. J. \& Fabricant, D. G. 1995, \aj, 110, 502

\bibitem[Dressler \& Shectman 1988]{dressler88} Dressler, A. \& Shectman,
S. A. 1988, \aj, 95, 985

\bibitem[Edge et al. 1990]{edg90} Edge, A. C., Stewart, G. C., Fabian, A. C., \&
Arnaud, K. A. 1990, \mnras, 245, 559

\bibitem[Edge, Stewart \& Fabian 1992]{edge92} Edge, A. C., Stewart, G. C., \&
Fabian, A. C. 1992, \mnras, 258, 177

\bibitem[Evrard 1988]{evrard88} Evrard, A. E. 1988, \mnras, 235, 911

\bibitem[Evrard 1990]{evrard90} Evrard, A. E. 1990, \apj, 363, 349

\bibitem[Evrard 1997]{evrard97} Evrard, A. E. 1997, \mnras, in press
(astro-ph/9701148)

\bibitem[Evrard, Metzler \& Navarro 1996]{evrard96} Evrard, A. E., Metzler,
C. A. \& Navarro, J. F. 1996, \apj, 469, 494

\bibitem[Fabian 1994]{fab94} Fabian, A. C. 1994, ARA\&A, 32, 277

\bibitem[Fabian et al. 1994]{fabian94} Fabian, A. C., Crawford, C. S.,
Edge, A. C., \& Mushotzky, R. F. 1994, \mnras, 267, 779

\bibitem[Fabricant et al. 1986]{fabricant86} Fabricant, D. G. Beers, T. C.,
Geller, M. J., Gorenstein, P., Huchra, J. P. \& Kurtz, M. J.
1986, \apj, 308, 530

\bibitem[Fabricant, Kent \& Kurtz 1989]{fabricant89} Fabricant, D. G.,
Kent, S. \& Kurtz, M. J. 1989, \apj, 336, 77

\bibitem[Geller \& Beers 1982]{geller82} Geller, M. J. \& Beers, T. C. 1982,
\pasp, 92, 421

\bibitem[Henriksen \& Markevitch 1996]{henrik96} Henriksen, M. J. \& Markevitch, M. L. 1996, \apj, 466, L79

\bibitem[Henry \& Briel 1995]{henry95} Henry, J. P. \& Briel, U. G. 1995, \apj, 443, L9

\bibitem[Hughes et al. 1993]{hughes93} Hughes, J. P., Butcher, J. A., Stewart, G. C. \& Tanaka, Y. 1993, \apj, 404, 611

\bibitem[Johnstone et al. 1992]{johnstone92} Johnstone, R. M., Fabian, A. C.,
Edge, A. C. \& Thomas, P. A. 1992, \mnras, 255, 431

\bibitem[Jones \& Forman 1984]{jones84} Jones, C. \& Forman, W. 1984, \apj,
276, 38

\bibitem[Jones \& Forman 1992]{jones92} Jones, C. \& Forman, W. 1992, in
{\it Clusters and Superclusters of Galaxies}, NATO ASI, VOl 366,
ed. A. C. Fabian (London: Kluwer), 49

\bibitem[J\o rgensen, Franx \& Kj\ae rgaard 1996]{jor96} J\o rgensen,
I. Franx, M.  \& Kj\ae rgaard, P. 1996, \mnras, 280, 167

\bibitem[Markevitch 1996]{marke96} Markevitch, M. 1996, \apj, 465, L1

\bibitem[Markevitch \& Vikhlinin 1997]{marke97} Markevitch, M. \& Vikhlinin, A. 1997, \apj, 474, 84

\bibitem[Matsuzawa et al. 1996]{matsu96} Matsuzawa, H., Matsuoka, M., Ikebe, Y., Mihara, T. \& Yamashita, K. 1996, \pasj, 48, 565

\bibitem[Metzler \& Evrard 1994]{metzler94} Metzler, C. A. \& Evrard, A. E.
1994, \apj, 437, 564

\bibitem[Metzler \& Evrard 1997]{metzler97} Metzler, C. A. \& Evrard, A. E.
1997, in preparation

\bibitem[Mitchell et al. 1979]{mitchell79} Mitchell, R. J.,
Dickens, R. J., Bell Burnell, S. J., \& Culhane, J. L. 1979, \mnras, 189, 329

\bibitem[Mohr et al. 1995]{mohr95} Mohr, J. J., Evrard, A. E., Fabricant, D. G.
\& Geller, M. J. 1995, \apj, 447, 8

\bibitem[Mohr, Geller \& Wegner 1996]{mohr96} Mohr, J. J., Geller, M. J. \&
Wegner, G. 1996, \aj, 112, 1816

\bibitem[Mohr \& Wegner 1997]{mohr97} Mohr, J. J. \& Wegner, G. 1997, \aj,
in press (astro-ph/9704202)

\bibitem[Mohr \& Evrard 1997]{mohr97b} Mohr, J. J. \& Evrard, A. E. 1997,
in preparation

\bibitem[Mushotzky \& Scharf 1997]{mushotzky97} Mushotzky, R. F. \&
Scharf, C. A. 1997, \apj, 482, L13

\bibitem[Navarro, Frenk \& White 1996]{navarro96} Navarro, J. F., Frenk, C. S.
\& White, S. D. M. \apj, 462, 563

\bibitem[Pildis, Bregman \& Evrard 1995]{pildis95} Pildis, R. A.,
Bregman, J. N. \& Evrard, A. E. 1995, \apj, 443, 514

\bibitem[Press et al. 1992]{press92} Press, W. H., Teukolsky, S. A., Vetterling, W. T. \& Flannery, B. P. 1992, Numerical Recipes in C, 2nd Edition,
(Cambridge University Press, Cambridge)

\bibitem[Roettiger, Burns \& Loken 1996]{rbl96} Roettiger, K.,  
Burns, J. O. \& Loken, C. 1996 \apj, 473, 651

\bibitem[Schindler 1996]{schindler96} Schindler, S. 1996, A\&A, 305, 756

\bibitem[Smith, Mushotzky \& Serlemitsos 1979]{smith79} Smith, B. W.,
Mushotzky, R. F., \& Serlemitsos, P. 1979, \apj, 227, 37

\bibitem[Snowden et al. 1994]{snow94} Snowden, S. L., McCammon, D., Burrows,
D. N. \& Mendenhall, J. A. 1994, \apj, 424, 714

\bibitem[Tamura et al. 1996]{tamura96} Tamura et al. 1996, \pasj, 48, 671

\bibitem[White et al. 1993]{white93} White, S. D. M., Navarro, J. F., Evrard, A. E.  \& Frenk, C. S. 1993, Nature, 366,429

\bibitem[Zabludoff \& Zaritsky 1995]{zabludoff95} Zabludoff, A. I. \& Zaritsky,
D. 1995, \apj, 447, L21

\end{thebibliography}
\end{document}